\documentclass[%
 reprint,
 amsmath,amssymb,
 aps,
prd,
]{revtex4-2}
\usepackage{amsmath}
\usepackage{amsfonts}
\usepackage{amssymb}
\newcommand{\be}{\begin{equation}}
\newcommand{\ee}{\end{equation}}

\usepackage{graphicx}
\usepackage{dcolumn}
\usepackage{bm}

\begin{document}


\title{A Non-Abelian Gauge Theory for Surface Excitations of $~^3$He-B}

\author{A. P. Balachandran}
 \email{apbal1938@gmail.com}
\affiliation{%
Physics Department, Syracuse University, \\
Syracuse, NY 13244-1130.
}%




\date{\today}

\begin{abstract}
In $^3$He-B, two atoms pair in an orbital angular momentum $1$ spin triplet state above the phase transition temperature with $SO(3) \times SO(3)$ symmetry. Below the transition temperature, this symmetry is spontaneously broken to the diagonal $SO(3)$ due to spin-orbit coupling. Considerations based on effective potentials and solitons show that $SO(3)$'s get enhanced to $SU(3)$'s and the symmetry 
breaking is that of $G= SU(3) \times SU(3)\times U(1)$ to $H= SU(3)$. The theory 
of the resultant Goldtsone modes can be naturally formulated as a gauge theory 
of $H$. Its Gauss law is treated here and shown to lead to surface states in a container with a dynamics governed by large gauge transformations.

  Observable consequences are pointed out. The transference of the analysis to the 
  chiral model of QCD is pointed out where $SU(3)$ are the left and right chiral 
  groups, $U(1)$ is the axial $U(1)$ , the surviving symmetry is flavour $SU(3)$ and the Goldstone modes are the pions.  

\end{abstract}

\maketitle


\section{\label{sec:level1}Introduction}

In $^3$He-B , the superfluid transition takes place by a pairing of this molecule in its $P$-wave spin triplet state with another such molecule in an identical state.

The order parameter for this transition can be denoted by a complex $3 \times 3$ matrix of fields $\phi=\{ \phi_{sm}(x)\}$ following Mermin \cite{Mermin1} where $SO(3)$ of spin
acts on $\phi$ by left multiplication and $SO(3)$ 
of orbital momentum acts on $\phi$ by right multiplication. Note, the indices $s,m$ take their values in the set $\{1,2,3\}$.

The chosen dynamics for $\phi$ is invariant under these
 actions of $SO(3)_L \times SO(3)_R$, $L$ and $R$ denoting the left and right groups.
 
 But this symmetry is spontaneusly broken to the diagonal 
 $SO(3)_J$ of total angular momentum by the boundary condition 
 \be 
 \lim_{| \vec{x}| \rightarrow \infty } \phi(x) = \mathbb{I}
 \ee 
Thus there is a spontaneous breakdown of the group $G= SO(3)_L \times SO(3)_R$ to the diagonal $H=SO(3)_J$. This breakdown is induced by the $L-S$ coupling.

This symmetry breaking pattern is reminiscent of the chiral symmetry group $SU(3)_L \times SU(3)_R$  of massless $QCD$ 
with three flavours and its spontaneous breakdown to the 
diagonal $SU(3)_{L+R}$. As alluded to above, we can in fact bring this resemblance closer by noting that $G$-invariant potentials 
are traces of polynomials in $\phi^\dagger \phi$ of the 
form 
\be 
\Sigma_n a_n \mathrm{Tr} ( \phi^\dagger \phi)^n .
\ee  
These potentials are invariant if $SO(3)_{L,R}$ is enlarged to $G= SU(3)_L \times SU(3)_R \times U(1)$ where $U(1)$ acts by an overall phase: $ \phi \rightarrow \omega \phi, | \omega| = 1 $. Then the vacuum symmetry group is $H= SU(3)_{L+R}$. These are our groups 
of interest.

The $U(1)$ group above is the analogue of the axial 
$U(1)$ of $QCD$. Just as in $QCD$, it is spontaneously broken here as well.

	$G$ does not act effectively on $\phi$. If $c=e^{\frac{2\pi i}{3}  \mathbb{I}_{3}} $, with $\mathbb{I}_3$ being the $3 \times 3 $ identity matrix of $SU(3)$,  generates the center of $SU(3)$, then the order $9$ discrete group $Z_9$ with elements diag$\{ c^m, \bar{c}^n\} \otimes \Omega$ with $n,m$ belonging to the set $\{1,2,3\}$ and $\Omega = e^{i (n-m) \mathbb{I}_{3}}$ acts trivially on $\phi$. Thus in quantum theory, only  representations of $G$ where $Z_9$ becomes identity are allowed.   With 
this proviso, we can work with $G$.

In Mermin's review , $\phi$ is restricted to $\omega \times R$ where $R$ is real and orthogonal. But a previous work \cite{Bal2} has found that for discussions about solitons of $~^3$He-B,  it is important to consider $\phi$ to be any $3 \times 3 $ matrix ( with the boundary condition  mentioned above). This paper will proceed in this latter direction. 

\section{\label{sec:level1}Gauge Theory of Goldstone Modes}

The next step in this paper is to describe the symmetry 
breakdown $G \rightarrow H$ as a gauge theory for $H$ for fields valued in $G$. This is an old result \cite{Bal_Gary_Al} reviewed below. What will be new here will involve the description of its superselection sectors following \cite{Gauss}, \cite{BalNairSasha}, \cite{Bal3} in sec. 3. This discussion involves new treatment of Gauss law and observables described in \cite{Gauss}.

We will see that for finite containers, they induce surface states and their dynamics as in ``topological insulators". Just as in the latter, their  origin is from spin-orbit coupling.\\

\textit{ Is $~^3$He-B a model in the lab for flavour physics of QCD?}
\\

Maybe so. It appears to have all the features of asymptotic QCD. But QCD has coloured fields like quarks.

What are their analogues here?

\section{\label{sec:level1}Spontaneous Breakdown as a Gauge Theory}

The paper of Balachandran, Stern and  Trahern \cite{Bal_Gary_Al} which treats this approach is summarised here. The discussion of the quantum generators of this gauge 
group, and its superselection sectors are reserved for the next section.

The gauge groups $\mathcal{G} , \mathcal{H}$ are the maps from spacetime to Lie groups $G,H$. In our considerations
, they are the restrictions of $\mathcal{G} , \mathcal{H}$ to constant time, say $0$ , that will be relevant. They are denoted by the same symbols. We consider four dimensional $M$ so that the constant time slices $M_3$ are three dimensional. It can be $\mathbb{R}^3$ or the 3-ball $B_3$ with coordinates $x$ and the boundary $S^2$. The evaluation of their elements at $x$ will be denoted by $g(x)$ and $h(x)$, respectively. Often we will suppress $x$.

The adaptations of this paper's arguments to any dimension is straightforward.

Let $\underline{G}$ and $\underline{H}$ denote the Lie algebras of $G$ and $H$ with basis \{$L_\lambda$\} and 
\{$T_\alpha$\}, in the defining representation of $G$ with 
$\mathrm{Tr} ( L_\lambda L_{\lambda'}) = 2 \delta_{\lambda \lambda'}$,$\mathrm{Tr} ( T_\alpha T_{\alpha'}) = 2 \delta_{\alpha \alpha'}$.

The adjoint action of $H$ leaves $\underline{H}$ invariant: 

\be
h T_\alpha h^{-1} = h_{\beta,\alpha} T_\beta 
\label{3.1}
\ee

where the matrix of this transformation in the basis 
\{$T(\alpha)$\} of $\underline{H}$ has also been denoted 
by $h$.

Hence it also leaves its orthogonal complement $\underline{G/H}$ in $\underline{G}$ invariant . If \{$S(i)$\} is its basis with  

 \be
 \mathrm{Tr} ~S(i) S(j) = 2 \delta_{ij} ,
 \ee
we can write

\be
h S(i) h^{-1} = [D(h)]_{ji} S(j)
\label{2.3}
\ee
where $D(h)$ is real and orthogonal. We sum on repeated indices here and below.

If $v$ is in $\underline{H}$,then 
\be
v = \frac{1}{2} T(\alpha) \mathrm{Tr} (v T(\alpha)),
\ee
while if $w$ is in its orthogonal complement ,

\be
w = \frac{1}{2} S(j) \mathrm{Tr} (w S(j)) .
\ee

The above equations are clearly valid if $h \in H$ is replaced by $h(x)$.

Now consider 
\be
A_\mu(g(x)) = \frac{i}{2} T(\alpha) \mathrm{~Tr~} T(\alpha)  g^{-1} (x) \partial_\mu g(x). 
\ee
Under the transformation 
\be
g(x) \rightarrow g(x) h(x), 
\ee
$A_\mu$ transforms as a connection for the gauge group $\mathcal{H}$:
\be
A_\mu(gh)= h^{-1} A_\mu h + i h^{-1} \partial_\mu h .
\label{3.6}
\ee
This follows from (\ref{2.3}) and the identity (\ref{3.6}) with $v_\mu= h^{-1} \partial_\mu h$. 

In contrast, 

\be
B_\mu(g) = \frac{i}{2} S(j) \mathrm{~Tr~} S(j)  g^{-1} (x)\partial_\mu g(x) 
\ee
transforms tensorially despite the derivative on $g$:
\be
B_\mu(gh) =  h^{-1} (x)B_\mu(g) h(x). 
\label{x1}
\ee

We can now consider $\mathcal{H}$-invariant interactions like 
\be
\mathrm{Tr~ } F_{\mu \nu} F^{\mu \nu} , \qquad \mathrm{Tr~} B_\mu B^\mu
\ee
 etc for the dynamics of the Goldstone modes, where $F$ is the field strength associated with the gauge connections $A$. But there are 
 more invariants available for $\mathcal{G, H}$ of $~^3$He-B . And there remains 
 the treatment of Gauss law which is in the next section.
 
 The gauge group $\mathcal{H}$ acts on $g_{L,R}$ according to $g_{L,R} \rightarrow g_{L,R} h $. Hence, 
 \be
 g_L g^\dagger_R \equiv \phi = \left\{ \phi_{s,i}\right\}
 \ee  
 is gauge invariant while  $s$ and $i$ carry $SU(3)_{L,R}$ actions. This is our order parameter.
 
 Also gauge connections  can be constructed for $\mathcal{G}_{L,R}$ separately giving two tensorial fields $B_{\mu~ L,R}$ and gauge connections $A_L, A_R$. So many sorts of interactions are 
 possible.

 \section{\label{sec:level1}Gauss Law and Superselection Rules}

 Let $A$ be a chosen connection. Following $QCD$, consider the action 
\be
S= -\frac{1}{4g^2}\int d^4x~ \mathrm{tr} F_{\mu \nu} F^{\mu \nu} + \int d^4x~  \mathrm{tr} A^\mu J_\mu
\label{4.1}
\ee
where $F_{\mu \nu}$ is the curvature of $A$ and $J^\mu$ is the covariantly conserved matter current.  The constants are as  in $QCD$.

In view of (\ref{x1}), $F$ transforms as $h^{-1} F h $ under an $\mathcal{H}$ gauge 
transformation so that $S$ is invariant even if $h$ is time dependent. 

So the generator of this gauge transformation must come from the 
Gauss law constraint.

The derivation of this constraint by quantising $g$ in the action is awkward. But we can get it simply by varying $A_0$ with an infinitesimal time dependent gauge transformation
\be
\delta g = g \epsilon, ~~\epsilon~ \textrm{time dependent}
\ee
where
\be
\epsilon = \epsilon^\alpha(x)\,  T(\alpha),
\ee
$\epsilon^\alpha$ being generic imaginary functions.

Since $D_i F_{i0}$ is $\underline{H}$ valued, and 
$\partial_0 \epsilon^\alpha$ are generic, we get the
Gauss law in the usual form :
\be
D_i F_{i0} \approx 0.
\label{gauss}
\ee

We can also add the matter coupling
\be
\int d^3x \mathrm{tr} ( A^\mu J_\mu )
\label{j1}
\ee
to the Lagrangian which will enhance (\ref{gauss}) by the matter term
$J_0$.


We must check that L.H.S of (\ref{gauss}) generates gauge transformations. and that requires the commutators of $A^i$ and $F_{0i}$. We can derive them 
as follows.

The analogue of the $p dq$ term can be read off from (\ref{4.1}) . It comes
from $\frac{1}{2g^2} \mathrm{tr } F_{0i} \partial_0 A_i$ from varying $A_i$ so that as in $QCD$, $A^i$ and $E_i = \frac{1}{2g^2} F_{0i}$ are conjugate. 

Then as in 
$QCD$ , the generator of small gauge transformations is 
\be 
Q_H  (X) = \int d^3 x ~\mathrm{tr }( D_i X E_i + X J_0 ),
\label{4.2}
\ee 
$X=$ smooth Lie-algebra valued compactly supported functions.
This is (\ref{gauss}) smeared with test functions  X.

Note that $E_i$ and $A^i$ are not gauge invariant.
They are not observables , but are elements of the field algebra.
 
 The classical Gauss law (\ref{gauss})   has the quantum version of `small ' gauge tranformations with generators $Q_H (X)$ which fulfill a) and b): 
 
 a) \be 
  Q_H (X) | ~\cdot~ \rangle = 0 
  \label{4.4}
  \ee
     {\it for all } such $X$ where  $| ~\cdot~ \rangle$
 are the allowed quantum vector states.

 b) If $K$ is an observable 
 \be \left[ Q_H(X), K \right] = 0.
 \label{4.5}
 \ee  

Then, $X$ being compactly supported , we recover classical Gauss law by partial 
integration , while (\ref{4.5}) is to ensure that observables preserve the domain
defined by (\ref{4.4}). ( There are technical issues with domains which remain 
to be answered.)

But one can also consider 
\be 
Q(\Xi) =  \int d^3 x~ \mathrm{Tr }( D_i \Xi  E_i + \Xi J_0 )
\ee
where $\Xi$ are smooth functions which may not vanish at infinity ( or the  boundary). These generate
 `large gauge transformations' or the Sky group \cite{sky}. As shown in previous papers, 
$Q(\Xi)$ commute with local observables due to local commutation relations , but in non-abelian 
theories such as $~^3$He-$B$ or $QCD$, they may not mutually commute. This has specific 
consequences for surface states as we shall see. 

 Two important points should be observed:
 
\begin{enumerate}
\item $Q(\Xi)$ generate a Lie algebra:
 \be
 \left[ Q(\Xi), Q(\Xi')\right] = i Q ([\Xi, \Xi']).  
 \ee
~~~But the rhs can be a small gauge transformation, so that the large ones by themselves do not generate a Lie algebra.
 
\item Adding a small gauge transformation $Q_H(X)$ to $Q(\Xi)$ is still a large gauge transformation $Q(\Xi + X)$ which has the same action on a quantum state in view of (\ref{4.4}): 
  \be 
  Q( \Xi + X ) | \cdot \rangle = \bigg(Q( \Xi) + Q(X )\bigg) | \cdot \rangle  = Q(\Xi) | \cdot \rangle 
  \label{4.7}
  \ee
  \end{enumerate}

 Further since $[ \Xi, X]$ is a compactly supported Lie algebra-function, 
   $Q([ \Xi, X ])$ is a small gauge transformation. ( As $X$ has a compact support, so does $[\Xi, X ]$.)
   
   So, let  ${\underline{\mathcal{H}}}$ be the Lie algebra of \textit{all} gauge transformations, large and small,  and $\underline{\mathcal{H}_{loc}}$ the Lie algebra of small ones. Then the latter is an invaraint subalgebra of the former and the large algebra is the quotient $\underline{\mathcal{H}} /\underline{ \mathcal{H}_{loc}}$. Its elements are equivalence classes $a + \underline{ \mathcal{H}_{loc}}$ where $a \in \underline{\mathcal{H}}$.
   
Generally the action of $Q(\Xi)$ on quantum states is determined by the asymptotic values $\Xi_\infty ( \hat{x}) $ of $\Xi$. That does not mean that we can replace $\Xi$ by its asymptotic function $\Xi_\infty$ in $Q(\Xi)$. That
 involves multiplying $\Xi$ by a delta-function supported at infinity leading
 to a function not smooth on spacetime. $Q(\Xi_\infty)$ is not an operator on Hilbert
  space. An example of such construction is electric charge, frequently written 
  as the integral $\oint_\infty \mathbf{E} \cdot d \mathbf{S}$ at infinity. It 
  may be acceptable in classical physics although its Poisson brackets make no
   sense. They are not operators on the quantum Hilbert space.
    
If $\psi$ is a gauge-noninvariant field such as a $~^3$He-B or a quark field
     of $QCD$, and elements of $\underline{\mathcal{H}}$  multiplies it, they are no longer
     gauge transformations , failing to fulfill requisite algebraic relations.

\subsection{Superselection Rules}

     Let $\mathcal{A}_{loc}$ denote the local algebra of gauge invariant observables. Its elements commute with all elements of the gauge algebra $\underline{\mathcal{H} }$.
      
    In a customary notation , $\underline{\mathcal{H} }= \mathcal{A}_{loc}'$. So in an irreducible representation of $\mathcal{A}_{loc}$, one can diagonalise a complete commuting set of operators (CCS) from $\underline{\mathcal{H}}$ and the center of its enveloping algebra. For $~^3$He-B confined in a ball, they are picked from $Q(\Xi)$ where $\Xi$ on the boundary $S^2$ of $\mathbb{R}^3$ or a spherical container of  $~^3$He-B are maps of $S^2$ to 
 $H = SU(3)_{L+R}$.

 We can illustrate these remarks by choosing the $8$ functions $\mathbb{I} \lambda_\alpha $ for $\Xi$ where $\mathbb{I}$ is the constant function with value $1$ on $\mathbb{R}^3$ or the ball and $\lambda_\alpha$ are the Gell-Mann matrices. Then, 
$Q_\alpha =Q( \mathbb{I} \lambda_\alpha) $ fulfill the $SU(3)$ Lie algebra
relation:
\be
[ Q_\alpha, Q_\beta ] = i f_{\alpha \beta \gamma} Q_\gamma .
\ee

We can diagonalise $ Q_{3,8}$ and
$Q_1^2 + Q_2^2 + Q_3 ^2 ( = I (I+1) )$  after fixing Casimirs to  values $c_2, c_3$ to obtain the vector

\be 
| \tilde{\lambda}_3, I , \tilde{\lambda}_8, c_2, c_3 \rangle = | 0 \rangle \otimes 
| \tilde{\lambda}_3, I , \tilde{\lambda}_8, c_2, c_3 \rangle ' .
\label{4.10} 
\ee
This vector is the analogue of the vacuum vector in quantum field theory.
On RHS, $|0\rangle$ is the Fock vacuum or the ground state and the primed vector carries just the $SU(3)$ representation.

The GNS construction then gives a Hilbert space representation of $\mathcal{A}_{loc}$.

Matter quantum numbers have been omitted in (\ref{4.10}).

Sticking to $SU(3)$, its Lie algebra has plenty of elements which do not commute  with $Q_{3,8}$, for example $Q_1 = Q(\mathbb{I} \lambda_1)$ if $\lambda_{~I} \neq 0 $. This will change the vector (\ref{4.10}) to a new one, but preserving $c_{2,3}$. As no element of $\mathcal{A}_{loc}$ 
can change $Q(\lambda_{3,8}\mathbb{I})$, this vector by GNS gives a new representation of $\mathcal{A}_{loc}$. 

The nonabelian symmetry $SU(3)_{L+R}$ is thus `spontaneously broken' to $U(1) \times U(1)$ with generators  $Q_{3,8}$ or to $U(2)$ if $I=0$ ( \cite{Gauss}, \cite{BalNairSasha}, \cite{Bal3}). 

These operators changing the diagonalised operators are also called `anomalous' . That is because they do not change the equations of motion ( which are  sensitive to only local values of $\Xi$ so that $Q(\Xi)$ act on them like small gauge transformations ) and so are classical symmetries, but change the representation of the local observables. 

Chiral `symmetry' of $QCD$ is an example.

\subsection{\label{sec:citeref}A Remark on Effective Lagrangians and Anomaly Matching}

Effective theories are subject to anomaly matching of 't Hooft \cite{thooft} so as to reproduce the anomalies of the parent theory.

The analysis outlined above shows that it is best to match the superselection algebras in the parent and effective theories. 

\subsection{Surface states in $^3$He-B}

 Consider $^3$He $B$ in say in a spherical ball . ( Containers with other
  topologies such as solid torus can also be treated). 
  
 The symplectic form which makes $A_i, E_j$ conjugate is valid for the action 
  $ \int F \wedge ^*F$  for potentials such as those of Mermin \cite{Mermin1} so that $Q_H(X)$ 
vanish on the quantum vector states $| ~\cdot~ \rangle = 0 $. This is the 
quantum Gauss law for the classical statement 
\be 
D_i E_i = 0 .
\ee              
        
Even in $QED$, this use of test functions is needed to express the generator 
of charge $U(1)$ as  pointed out earlier. 

Recalling that $E_i$ are operator valued distributions, such smearing 
gives us operators on a Hilbert space and is also suggested by Sobolev's treatment 
of elliptic operators on manifolds with boundaries. 

But then $Q(\Xi)$ need not vanish on $|~ \cdot~ \rangle$ used in the 
GNS construction. It generates a Lie algebra and a complete 
commuting set (CCS) on diagonalisation gives a vector state 
$| \tilde{\lambda}_3, I , \tilde{\lambda}_8, c_2, c_3 \rangle $ suitable for creating a
 representation of $\mathcal{A}_{loc}$. ( We have suppressed the Fock state 
 and diagonalised matter quantum numbers.)
 
 The field equations $F(\psi)=0$ involve only local fields $\psi$
 and are gauge invariant:
 \be
 F(\psi)=0 \Rightarrow F( [ Q(\Xi), \psi])=0
 \label{4.11}
  \ee    
          
As $\Xi$ can be an $X$, (\ref{4.11}) implies full gauge invariance 
even though $\psi$ may not commute with $Q(\Xi)$ as is the case with connections.

Such a situation is familiar from $QED$ where the `chemical potential' 
$N$ generates $U(1)$ transformations.

Adding $N$ to the $QED$ Hamiltonian $H_0 $, we work with the new Hamiltonian 
$H= H_0 + N$ which gives exactly the same equations of motion, but 
affects the partition functions. 

For such reasons , let us call $Q(\Xi)$ as generalised chemical potentials. If $h_0$ 
is the $^3$He-B Hamiltonian, we start with the `vacuum' 
\be
|0 \rangle,  \qquad  Q(\Xi_i) |0 \rangle =0 , 
\ee
then we can go to the sector with $SU(3)_{L+R}$ quantum numbers as in (\ref{4.10}). This 
is the analogue of the excited states considered by Doplicher, Haag and Roberts \cite{DHR}. 
Expectation values of local operators change on the
folium of states (\ref{4.10}) . An example is given below. 

Let $\mu$ be a constant with dimension of energy. We can then introduce a generalized chemical potential 
in a new Hamiltonian 
\be
h = h_0 + \mu Q(\Xi)
\label{4.12}
\ee
where $Q(\Xi)$ does not commute with the $Q_{3,8}$. Then the vector state 
(\ref{4.10}) will change in time and so will the expectation of the generic  surface operators 
$Q(\Xi')$. Thus one has dynamics on the boundary two sphere $S^2$. 

For example,  suppose $Q(\Xi)= a Q_1 + b Q_5$ with $a/b$ being irrational. The time $t=0$ 
inital state (\ref{4.10}) changes with time and is not periodic since $\exp( i t\mu Q(\Xi))$ 
has no period. One shows that it keeps evolving the diagonalised operators from one to 
another Cartan subalgebra \cite{Bal3} .  

 One can prove that it never returns to the
 initial state of $t=0$. Its orbit is ergodic, being dense in the Grassmannian $Gr_2 (C^3)$ .

The spectrum of $h$ will keep changing with time. If $\mu$ is chosen to be time-dependent and is switched 
off at time $T$, there will be transitions to the ground state. Maybe they can be seen. 

The surface operators $Q(\Xi)$ emerge from spin-orbit coupling as 
seen in the literature on topological insulators. But why are they `topological' ? 
Perhaps because  $Q(\Xi)$ has no spacetime metric in it. But they do have a Lie algebra metric.

\subsection{Interpretation of the Generalised Chemical Potential}

In $QED$, the term $N$ comes from coupling the system to an external source supplying the charge.
 The interpretation of $\mu Q(\Xi)$ is much the same, it comes from the measuring apparatus 
 coupled to the system. In a paper under preparation, this claim is confirmed using Takesaki duality and a measurement theory for guage fields is outlined.

Expectation values of local observables change when the superselection 
sector is changed. This happens already in $QED$. 

Expectation value of charge density $:J_0:$ vanishes in the vacuum sector,
 but it cannot identically vanish if  $N$ , its integral, has the value $q$ in the charge $q$   sector. In this case, 
 we have to construct $: J_0 :$ from charged fields
like the electron field. The last is not gauge invariant. It is associated 
with the field algebra.

There are similar gauge invariant fields in any local gauge theory 
such as that of $^3$He-$B$. The operators $\mathrm{tr } (E^2 (x)+ B^2(x))
 , \mathrm{tr } E \cdot B $ are examples. Combination of such terms
integrated gives the Hamiltonian in the absence of chemical potential
and its expectation value keeps changing in time when the chemical 
potential keeps changing the vector state. Knowing their expectation 
values at time $t=0$ when the experiment starts and the chemical 
potential $\mu(t) Q(\Xi)$  is switched on ( it is now time dependent ) 
and at time $T$, when  it is switched off. one infers the properties of the local dynamics. 

\subsection{An Important Remark }

The chemical potential $Q(\Xi)$ is intrinsically present in the Hamiltonian 
emergent from the action 
\be
S= \int d^4x~ \mathrm{tr}~ (\frac{1}{2e^2}( \mathbf{E}^2 - \mathbf{B}^2) +  A^\mu J_\mu ).
\label{4.13}
\ee

The exact Hamiltonian is 
\be 
\int d^3x~ \mathrm{tr~} \frac{1}{2e^2}( \mathbf{E}^2 + \mathbf{B}^2) + Q(A_0).
\ee

 The standard treatments set $A_0 = 0 $, but one sees that it plays 
 the role of $\mu Q(\Xi)$.  Its asymptotic value seems determined by experimental setup.  Having 
 fixed the vector state, its evolution is also governed by the chemical potential $Q(A_0)$.
 
 In $QED$, $A_0$ is the electrostatic potential . By fluctuating it, the total energy 
 gets fluctuated and causes observable transitions.
 
 Likewise, the chemical potential $\mu Q(\Xi)$ should cause observable
 effects and help in probing the system.

\section{Final Remarks}

This paper is restricted to considering surface effects. Still it suggests 
observable effects such as energy fluctuations and transitions due to them from emission of 
probably Goldstone modes. 

~~In addition, the approach predicts solitons \cite{Bal2}. 
It predicts as well theta vacua which are superselection sectors which can be 
constructed for $~^3$He-B following \cite{BalNairSasha} and which are spinorial. 
Such matters are being omitted 
for brevity. 

Superselection sectors like the charged vectors in $QED$ can be constructed using fields 
$\psi$ ( or the Cuntz algebra following \cite{DHR}). They will be needed for detailed calculations.

The dynamics on the surface are determined by $Q(\Xi)$. The functions 
$\Xi$ can be $\Xi^\alpha ( \hat{x}) \lambda_\alpha$ on the surface 
where $Q(\Xi)$ dominate the dynamics. If $\Xi^\alpha ( \hat{x})$ are constants on $S^2$ 
these being the case briefly considered above, the surface dynamics may be 
soluble. They seem related to the quantum KAM dynamics treated in \cite{Facchi}.
 
 In that connection, note that in the example above with irrational $a/b$ in  $ a Q_1 + b Q_5$ , 
 the evolution under conjugation cannot bring back $Q_{3,8}$ to the same Cartan subalgebra, as 
 only automorphisms of the Cartan subalgebra are the elements of the Weyl group, which is $S_3$ 
 where the elements have periods $2,3$ or 6. So if $\exp(i \tau ( a Q_1 + b Q_5))$ maps back 
 $Q_{3,8}$ to the same Cartan subalgebra, one of these periods has to be  an integer multiple of 
 $6 \tau$. But that  is not possible.
 
 The considerations of this paper are valid for the $QCD$ chiral model on $R^3$ ( or $S^3$) . 
 They remain to be studied. 

 \begin{acknowledgments}
 This paper could not have been  written without the help of Arshad Momen and Babar Qureshi. I thank them. Helpful comments from Arshad, Parameswaran Nair, Al Stern and Rakesh Tibrewala are also appreciated. I also thank 
 the Institute of Mathematical Sciences,Chennai and the hospitality of my colleagues there ( especially the Director Ravindran and S. Digal) where this work was completed. 
\end{acknowledgments}

%

\providecommand{\noopsort}[1]{}\providecommand{\singleletter}[1]{#1}%

\end{document}